\documentclass[10pt]{elsart}

\date{}

\usepackage{amsmath}
\usepackage{amssymb}
\usepackage{makeidx}
\usepackage{graphicx}
\theoremstyle{plain}
\newtheorem{theo}{Theorem}
\newtheorem{corr}{Corollary}
\newtheorem{lema}{Lemma}
\newtheorem{propp}{Proposition}
\newtheorem{conjj}{Conjecture}

\usepackage{cite}
\usepackage{color}
\providecommand{\com[2]}{\begin{tt}[#1: #2]\end{tt}}

\newcommand{\rj}[1]{{#1}}

\newcommand{\vecte}{{\bf 1}}

\newcommand{\re}{{\mathbb R}}
\newcommand{\n}{{\mathbb N}}

\newcommand{\jsr}{{\hat \rho}}
\newcommand{\lsr}{{\check \rho}}

\include{references.bib}

\begin{document}

\begin{frontmatter}

\title{Overlap-free words and spectra of matrices}

\author{
Rapha\"el M. Jungers}
\address{Division of Applied Mathematics,
Universit\'e catholique de Louvain, 4 avenue Georges Lemaitre,
B-1348 Louvain-la-Neuve, Belgium, Email: raphael.jungers@uclouvain.be}

\author{
Vladimir Yu.~Protasov}
\address{Department of Mechanics and
Mathematics, Moscow State University, Vorobyovy Gory, Moscow,
119992, Russia, Email: v-protassov@yandex.ru}

\author{
Vincent D. Blondel}
\address{Division of Applied Mathematics,
Universit\'e catholique de Louvain, 4 avenue Georges Lemaitre,
B-1348 Louvain-la-Neuve, Belgium, Email: vincent.blondel@uclouvain.be}

\begin{abstract}
Overlap-free words are words over the binary alphabet $A=\{a, b\}$
that do not contain factors of the form $xvxvx$, where $x \in A$
and $v \in A^*$. We analyze the asymptotic growth of the number
$u_n$ of overlap-free words of length $n$ as $\, n \to \infty$. We
obtain explicit formulas for the minimal and maximal rates of
growth of $u_n$ in terms of spectral characteristics  (the lower
spectral radius and the joint spectral radius) of certain  sets of
matrices of dimension $20 \times 20$. Using these descriptions we
provide new estimates of the rates of growth that are within $0.4\%$ and $0.03 \%$ of their exact values. The
best previously known bounds were within $11\%$ and $3\%$
respectively. We then prove that the value of $u_n$ actually has
the same rate of growth for ``almost all'' natural numbers $n$.
This ``average'' growth is distinct from the maximal and minimal
rates and can also be expressed in terms of a spectral quantity
(the Lyapunov exponent). We use this expression to estimate it. In
order to obtain our estimates, we introduce new algorithms to
compute spectral characteristics of sets
of matrices. These algorithms can be used in other contexts and are of independent interest.%

\end{abstract}


\end{frontmatter}

\section{Introduction}

\begin{figure}
\centering
\begin{tabular}{cc}
\includegraphics[scale = .4]{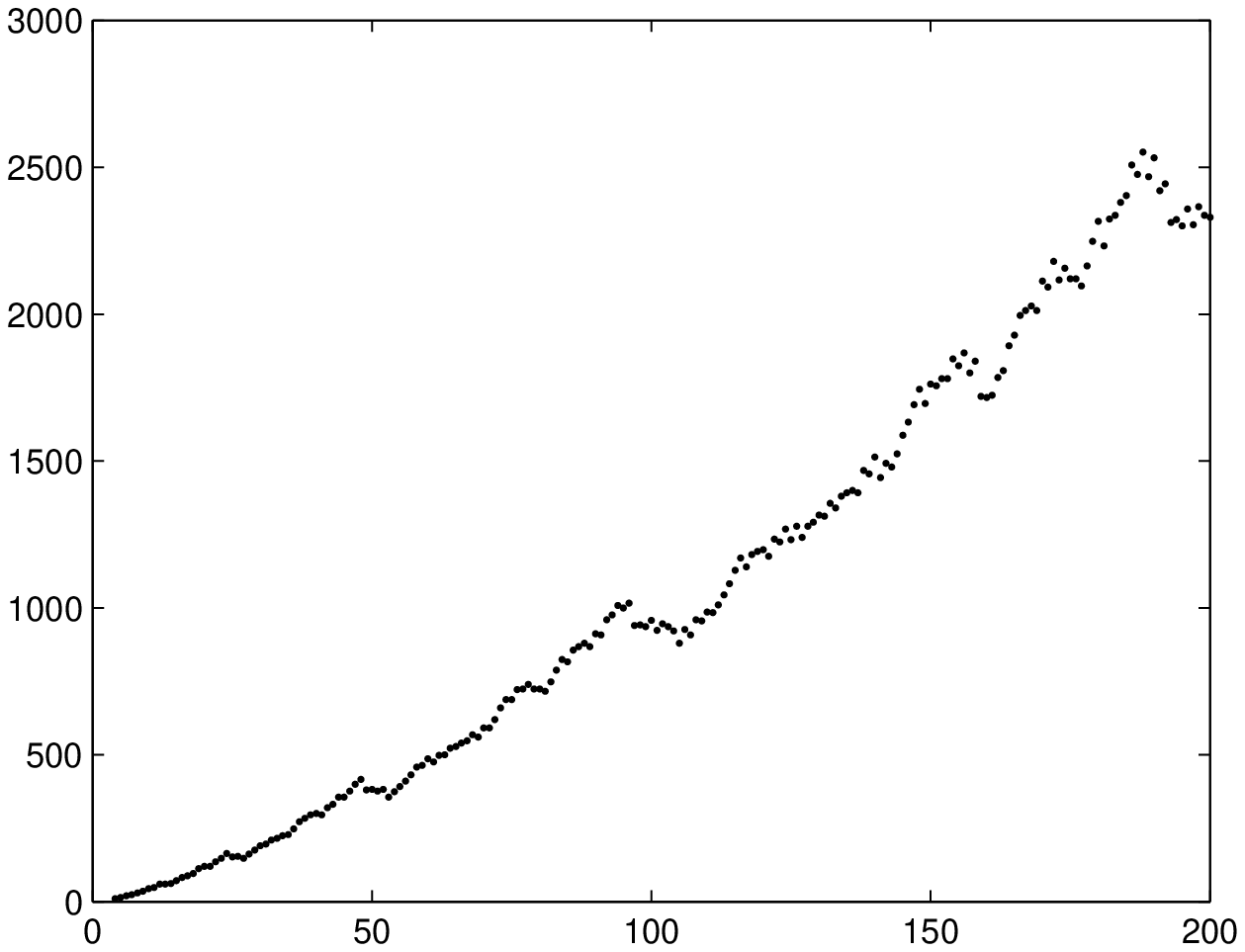}&

\includegraphics[scale = .4]{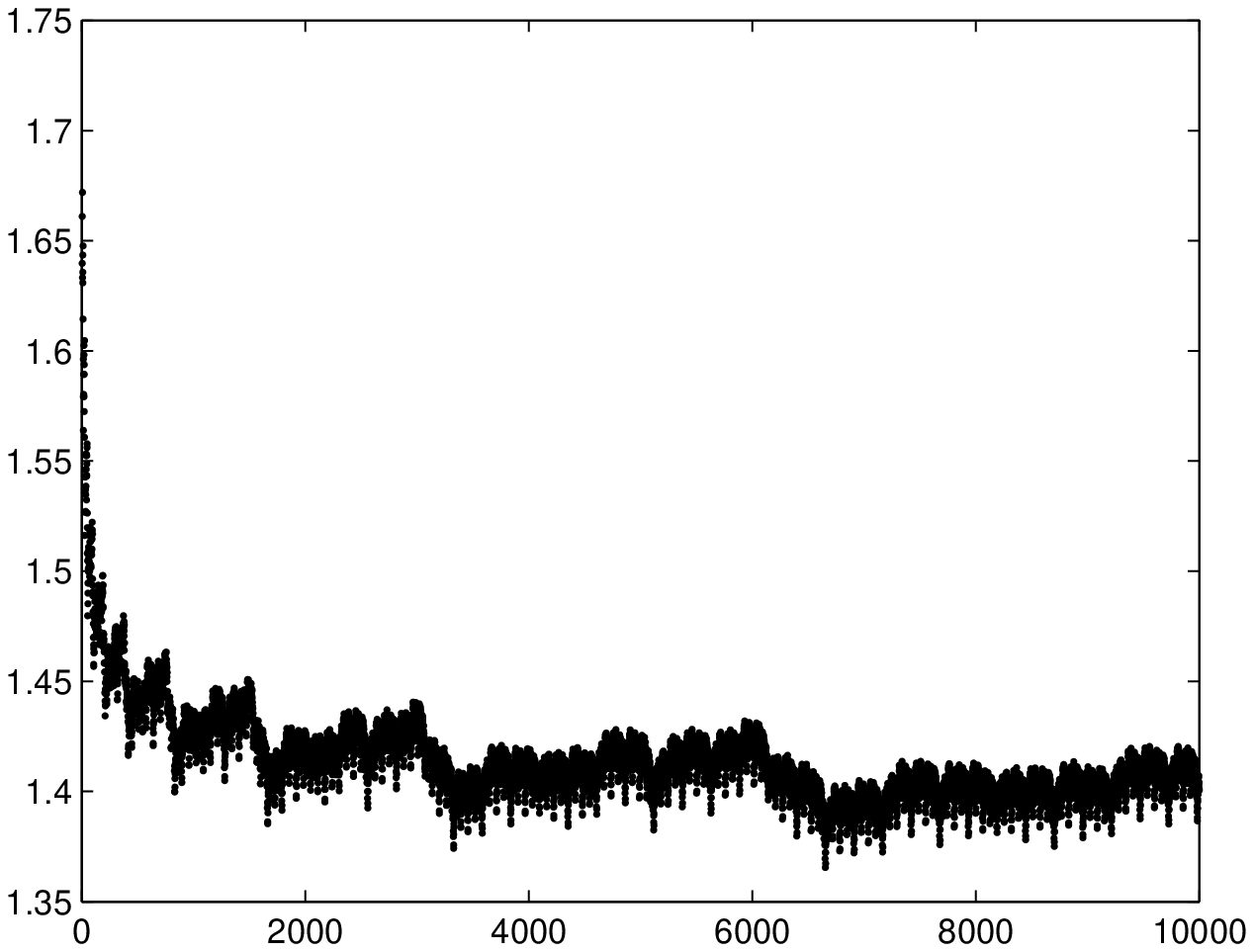}\\
(a)&(b)
\end{tabular}
\caption{The values of $u_n$ for $1\leq n \leq 200$ (a) and $\log{u_n}/\log{n}$ for $1\leq n \leq 10000$
(b).}\label{fig-un}
\end{figure}

Binary overlap-free words have been studied for
more than a century.  These are words over
the binary alphabet $A=\{a,b\}$ that do not contain factors of
the form $ xvxvx, $ where $x\in A$ and $v\in A^*.$  For instance, the word \emph{baabaa}
is overlap free, but the word \emph{baabaab} is not, since it can be written $xuxux$ with $x=b$ and $u=aa.$  See \cite{berstel-repetition} for a recent survey.  Thue \cite{thue1,thue2} proved in 1906 that there are
infinitely many overlap-free words.  Indeed, the well-known
Thue-Morse sequence\footnote{The Thue-Morse sequence is the
infinite word obtained as the limit of $\theta^n(a)$ for $n
\rightarrow \infty$ with $\theta(a)=ab,\ \theta(b)=ba;$ see
\cite{cassaigne}.} is overlap-free, and so the set of its factors
provides an infinite number of different overlap-free words. The
asymptotics of the number $u_n$ of such words
  of a given length $n$ was analyzed in a number of subsequent contributions\footnote{The number of
  overlap-free words of length $n$  is referenced in
the On-Line Encyclopedia of Integer Sequences under the code A007777; see  \cite{online-enc}.
The sequence starts 1, 2, 4, 6, 10, 14, 20, 24, 30, 36, 44, 48, 60, 60, 62, 72,...}. The number of
factors of length $n$ in the
Thue-Morse sequence is proved in \cite{brlek} to be larger than $3n$, thus providing a
linear lower bound on $u_n$:
$$
u_n\ \geq \ 3\, n.
$$
The next improvement was obtained by Restivo and Salemi
\cite{restivosalemi}. By using a certain decomposition result,
they showed that the number of overlap-free words grows at most
polynomially:
$$
u_n\ \leq \ C \, n^{\, r},
$$
where $r=\log(15) \approx 3.906.$  This bound has been sharpened
successively by Kfoury~\cite{kfoury}, Kobayashi~\cite{koba}, and
finally by Lepisto~\cite{lepisto} to the value $r=1.37$. One could
then suspect that the sequence $u_n$ grows linearly.  However,
Kobayashi~\cite{koba} proved that this is not the case.  By enumerating
the subset of overlap-free words of length $n$ that can be infinitely extended to the right
he showed that $ u_n \ \ge \  C\, n^{\, 1.155}$
and so we have
$$ C_1 \, n^{\, 1.155} \leq u_n \leq  C_2\, n^{\, 1.37}.$$
It is worth noting that the sequence $u_n$ is $2$-regular, as shown by Carpi \cite{carpi}.  On Figure \ref{fig-un}(a) we show the values of the sequence $u_n$ for $1\leq n \leq 200$ and on
Figure \ref{fig-un}(b) we show the behavior of $\log{u_n}/\log{n}$ for larger values of $n$.
One can see that the sequence $u_n$ is not monotonic, but is globally increasing with $n$.
Moreover the sequence does not appear to have a polynomial growth since the value $\log{u_n}/\log{n}$
does not seem to converge.
In view of this, a natural question arises: is the sequence
$u_n$ asymptotically equivalent to $n^r$ for some $r$~?
Cassaigne proved in~\cite{cassaigne} that the answer is negative.
He introduced the lower and the upper exponents of growth:
\begin{eqnarray}\label{alphabeta}
\alpha & = & \sup  \bigl\{ r \ \bigl|  \exists  C > 0,  u_n  \ge  C n^r  \bigr\},\\
\nonumber \beta & = & \inf  \bigl\{ r \ \bigl|  \exists  C > 0,  u_n  \le  C n^r  \bigr\},
\end{eqnarray} 
and showed that $\, \alpha < \beta$. Cassaigne made a real breakthrough in the study of overlap-free
words by characterizing in a
constructive way the whole set of overlap-free words. By improving
the decomposition theorem of Restivo and Salemi he showed that
the numbers~$u_n$ can be computed as sums of variables that are
obtained by certain linear recurrence relations. These relations are explicitly given in the next section
and all
numerical values can be found in Appendix \ref{ap-values}. As a result of this description, the number of
overlap-free words of length $n$ can be
computed in logarithmic time. For the exponents of growth Cassaigne has also
obtained the following bounds: $\alpha<1.276$ and $\beta> 1.332.$
Thus, combining this with the earlier results described above, one has the following inequalities:
\begin{equation}\label{previous}
1.155 \ <\ \alpha \ <\ 1.276 \qquad \mbox{and} \qquad  1.332 \ <\ \beta\ <\ 1.37.
\end{equation}

In this paper we develop a linear algebraic approach to study the
asymptotic behavior of the number of overlap-free words of
length~$n$. Using the results of Cassaigne we show in
Theorem~\ref{th1} that $u_n$ is asymptotically equivalent to the
norm of a long product of two particular matrices $A_0$ and $A_1$ of
dimension $20 \times 20$. This product corresponds  to the binary
expansion of the number $n-1$.  Using this result we express the
 values of $\alpha$ and $\beta$ by means of certain joint spectral
characteristics of these matrices. We prove that $\, \alpha\, =\,
\log_2 \check \rho(A_0,A_1)\, $ and $\, \beta \, = \, \log_2
\jsr(A_0,A_1)$, where $\lsr$ and
$\jsr$ denote, respectively, the  lower
spectral radius and the joint spectral radius of the matrices
$A_0,A_1 $ (we define these notions in the next section). In Section~\ref{section:numerical}, we
estimate these values and we obtain the following improved bounds
for $\alpha$ and $\beta$:
\begin{equation}\label{ours}
1.2690 \ <\ \alpha \ <\ 1.2736   \qquad \mbox{and} \qquad  1.3322 \ <\ \beta\ <\ 1.3326.
\end{equation}
Our estimates  are, respectively, within $0.4 \%$ and $0.03 \%$ of the exact values. In addition, we show
in Theorem~\ref{th2} that the smallest and the largest rates of growth
of $u_n$ are effectively attained, and there exist positive constants
$C_1, C_2$ such that $\ C_1 \, n^{\alpha }\ \le \ u_n \ \le \ C_2\, n^{\, \beta}$ for all $n \in \n$.

Although the sequence $\, u_n\, $ does not exhibit an asymptotic
polynomial growth, we then show in Theorem~\ref{th5} that for
``almost all'' values of $n$ the rate of growth is actually equal
to \ $\, \sigma \, = \, \log_2 \, \bar \rho (A_0, A_1)$, where
$\bar \rho$ is the Lyapunov exponent of the matrices. For almost
all values of $n$ the number of overlap-free words does not grow
as $n^\alpha$, nor as $n^\beta$, but in an intermediary way, as
$n^\sigma$. This means in particular that the value $\frac{\, \log
\, u_n}{\log \, n}\, $ converges to $\sigma$ as $\, n \to \infty$
along a subset of density $1$. We obtain the following bounds for
the limit $\sigma,$ which provides an estimation within $0.8 \%$
of the exact value:
$$
1.3005  < \sigma <  1.3098.
$$
These bounds clearly show that $\, \alpha < \sigma < \beta$.\\

To compute the exponents $\alpha$ and $\sigma$ we introduce new
efficient algorithms for estimating the lower spectral radius
$\check \rho$ and the Lyapunov exponent $\bar \rho$ of {matrices}.
These algorithms are both of independent interest as they can be
applied
to arbitrary matrices.\\

Our linear algebraic approach
not only allows us to
improve the estimates of the asymptotics of the
number of overlap-free words, but also clarifies some aspects of the nature
of these words. For instance, we show that the ``non purely
overlap-free words'' used in~\cite{cassaigne} to compute $u_n$ are asymptotically negligible when
considering the total number of overlap-free words.\\

The paper is organized as follows. In the next section we
formulate and prove the main theorems (except for
Theorem~\ref{th1}, whose proof is quite technical and is given in
Appendix \ref{ap-th1}). Then in Section~\ref{section:numerical} we
present algorithms for estimating the joint spectral radius, the
lower spectral radius, and the Lyapunov exponent of linear
operators. Applying them to those special  matrices we obtain the
estimations for $\, \alpha , \beta $ and $\sigma$. In the
appendices we write explicit forms of the matrices and initial
vectors used to compute $u_n$, we give a proof of
Theorem~\ref{th1} and present the results of our numerical
algorithms.

\section{The asymptotics of the overlap-free words}\label{section:theoretical}
In the sequel we use the following notation: $\re^d$ is the
$d$-dimensional space, inequalities $x \ge 0$ and $A \ge 0$ mean
that all the entries of \rj{the} vector $x$ (respectively, of the
matrix $A$) are nonnegative. We denote $\re^d_+ = \{x \in \re^d,\
x \ge 0\}$, by $|x|$ we denote a norm of the vector $x \in \re^d$,
and by  $\| \cdot
 \|$ any matrix norm. In particular, $
|x|_1  =  \sum\limits_{i=1}^d|x_i|,\ \|A\|_1  =
\sup_{|x|_1=1}|Ax|  =  \max_{j = 1, \ldots  d}
\sum_{i=1}^d|A_{ij}|$. We write $\vecte$ for the vector $(1,
\ldots , 1)^T\in \re^d$, $\rho(A)$ for the spectral radius of the
matrix $A$, that is, the largest magnitude of its eigenvalues. If
$A \ge 0$, then there is a vector $v \ge 0$ such that $ Av =
\rho(A)v$ (the so-called Perron-Frobenius eigenvector).  For two
functions $f_1, f_2$ from a set $Y$ to $\re_+$ the relation $
f_1(y) \asymp f_2(y) $ means that there are positive constants $
C_1, C_2$ such that $ C_1 f_1(y) \le
f_2(y) \le  C_2 f_1(y)$ for all $ y \in Y$.\\
To compute the number $ u_n $ of overlap-free words of length
$n$ we use several results from \cite{cassaigne} that we summarize in the following theorem:\\
\begin{theo}\label{cassaigne}
Let $F_0,F_1\in \re^{30\times 30},$ and let $w,y_3,\dots
,y_{15}\in \re_+^{30}$ be as given in Appendix  \ref{ap-values}.
For $n\geq 16$, let $y_n$ be the solution of the following
recurrence equations
\begin{equation} \label{eq-cassaigne}
\begin{array}{ccc}
y_{2n}&=&F_0y_n\\
y_{2n+1}&=&F_1y_n.
\end{array}
\end{equation}
Then, for any $n\geq 16,$ the number of overlap-free words of length $n$ is equal to $w^Ty_{n-1}.$
\end{theo}
It follows from this result that the number $u_n$ of overlap-free words of length $n\geq 16$ can be
obtained by first computing the binary expansion $d_k \cdots d_1$ of $n-1$, i.e.,
$n-1 = \sum_{j=0}^{k-1}d_{j+1}2^{j}$, and then defining
\begin{equation}\label{products}
u_{n}   =  w^T F_{d_1}\cdots F_{d_{k-4}}y_{m}
\end{equation}
where  $m =d_{k-3}+d_{k-2} 2 + d_{k-1}2^2 +  d_k2^3$.
To arrive at the results summarized in Theorem 1, Cassaigne builds a system of recurrence equations
allowing the computation of a vector $U_n$ whose entries are the number of overlap-free words of certain
types (there are $16$ different types).  These recurrence equations also involve the recursive computation
of a vector $V_n$ that counts other words of length $n,$ the so-called ``single overlaps".
The single overlap words are not overlap-free, but have to be computed, as they generate overlap-free
words of larger lengths.
 We now present the main result of this section which improves the above theorem in two directions.
 First we reduce the dimension of the matrices from 30 to 20, and second we prove that $u_n$ is given
 asymptotically by the norm of a matrix product.  The reduction of the dimension to 20 has a
 straightforward interpretation: when computing the asymptotic growth of the number of overlap-free words,
 one can neglect the number of ``single overlaps" $V_n$ defined by Cassaigne.  We call the remaining words
 \emph{purely overlap-free words,} as they can be entirely decomposed in a sequence of overlap-free
 words via Cassaigne's decomposition (see \cite{cassaigne} for more details).\\

\begin{theo}\label{th1}
Let $A_0,A_1\in \re_+^{20 \times 20}$ be the matrices defined in Appendix A (Equation \ref{A0A1}),
let $ \|\cdot \| $ be a matrix norm, and let $A(n):\n\rightarrow \re_+^{20\times 20}$ be defined as
$A(n)=A_{d_1}\cdots A_{d_k}$ with $d_k\dots d_1$ the binary expansion of $n-1$. Then,
\begin{equation}\label{express}
u_n \asymp ||A(n)||.  
\end{equation}
\end{theo}
Observe that the matrices $F_0,F_1$ in Theorem \ref{cassaigne} are both nonnegative and hence possess
a common invariant
cone $K = \re^{30}_+$. We say that a cone $K$ is invariant for a
linear operator $B$ if $BK \subset K$. All cones are assumed to be
solid, convex, closed, and pointed. We start with the following simple
result proved in \cite{protasov3}.
\begin{lema}\label{l1}
For any cone $K \subset \re^d$, for any norm $| \cdot  |$ in
$\re^d$ and any matrix norm $\| \cdot  \|$ there is a
homogeneous continuous function $\gamma : K \to \re_+$ positive on
${\rm int} K$ such that for any $x \in {\rm int} K$ and for
any  matrix $B$ \rj{that leaves $K$ invariant} one has
$$
\gamma(x) \|B\| \cdot |x|  \le  |Bx|  \le
\frac{1}{\gamma(x)} \|B\| \cdot |x|.
$$
\end{lema}
\begin{corr}\label{c1}
Let two matrices $A_0, A_1$ possess an invariant cone $K
\subset \re^d$. Then for any  $x \in {\rm int} K$ we have
$|A_{d_1} \cdots A_{d_k}x|  \asymp  \|A_{d_1} \cdots
A_{d_k}\|$ for all $k$ and for all indices $d_1, \ldots , d_k \in
\{0,1\}$.
\end{corr}
In view of Corollary~\ref{c1} and of Eq. (\ref{products}), Theorem~\ref{th1} may seem obvious, at
least if we consider the matrices $F_i$ instead of $A_i$. One can however not directly apply Lemma~\ref{l1}
and Corollary~\ref{c1} to the matrices $A_0, A_1$ or to the matrices $F_0, F_1$ because the vector
corresponding to $x$ is not in the interior of the positive orthant, which is an invariant cone of these
matrices.
To prove Theorem~\ref{th1} we construct a wider invariant cone of
$A_0$ and $A_1$ by using special properties of these matrices. We
detail the construction of this cone in the proof given in
Appendix \ref{ap-th1}. Theorem~\ref{th1} allows us to express the
rates of growth of the sequence $u_n$ in terms of norms of
products of the matrices $A_0, A_1$ and then to use joint spectral
characteristics of these matrices to estimate the rates of growth.
More explicitly, Theorem \ref{th1} yields the
following corollary:
\begin{corr}\label{p1}
Let $A_0,A_1\in \re_+^{20 \times 20}$ be the matrices defined in Appendix A and let
$A(n):\n\rightarrow \re_+^{20\times 20}$ be defined as $A(n)=A_{d_1}\cdots A_{d_k}$ with
$d_k\dots d_1$ the binary expansion of $n-1$. Then
\begin{equation}\label{e1}
\frac{\log u_n}{\log n} - \log  \|A(n)\|^{1/k}\to \ 0 \quad \mbox{as}\quad n \to \infty .
\end{equation}
\end{corr}
{\it Proof.  } Since $ \frac{\log_2 n}{k}  \to  1 $ as $ n
\to \infty$, we have
$$\begin{tabular}{rll}
$\lim\limits_{n \to \infty} \left(\frac{\log_2 u_n}{\log_2 n}
 -  \frac{\log_2  \|A_{d_1}\cdots A_{d_k}\|}{k}\right)$&$=$&\\
$\lim\limits_{n \to \infty} \frac{\log_2 u_n - \log_2
\|A_{d_1}\cdots A_{d_k}\|}{k}$&$=$&$\lim\limits_{n \to \infty} \frac{\log_2 \bigl(
u_n  \cdot \|A_{d_1}\cdots A_{d_k}\|^{-1}\bigr)}{k}.$
\end{tabular}$$
By Theorem~\ref{th1} the value $\log_2 \bigl( u_n  \cdot
\|A_{d_1}\cdots A_{d_k}\|^{-1}\bigr)$ is bounded uniformly over $n
\in \n$, hence it tends to zero, being divided by $k$.{\hfill $\Box$}\\
We first analyze the smallest and the largest exponents of growth
$\alpha$ and $\beta$ defined in Eq. (\ref{alphabeta}). For a given
set of matrices $\Sigma = \{A_1, \ldots, A_m\}$ we denote by
$\lsr$ and $\jsr$ its lower spectral radius and its joint spectral
radius:
\begin{eqnarray}\label{jsrlsr}
\lsr (\Sigma)&=& \lim_{k \to
\infty}\min_{d_1, \ldots , d_k  \in  \{1, \ldots , m\}}
\|A_{d_1}\cdots A_{d_k}\|^{1/k},\\
\nonumber \jsr (\Sigma)&=& \lim_{k \to \infty}\max_{d_1, \ldots ,
d_k  \in  \{1, \ldots , m\}} \|A_{d_1}\cdots A_{d_k}\|^{1/k}. \end{eqnarray}
Both limits are well-defined and do not depend on the chosen norm.
Moreover, for any product $A_{d_1}\cdots A_{d_k}$ we have
\begin{equation}\label{rho}
\lsr \le  \rho(A_{d_1}\cdots A_{d_k})
^{1/k}\le \jsr
\end{equation} (see \cite{protasov1,blondel98survey} for  surveys on these notions).
\begin{theo}\label{th2}
For $ k
\ge 1 $, let $\alpha_k   =  \min\limits_{2^{k-1}<
n \le 2^k} \frac{\log u_n}{\log n}$ and $\beta_k   =
 \max\limits_{2^{k-1}< n \le 2^k}\frac{\log u_n}{\log n}$. Then
\begin{equation}\label{ab}
\alpha = \lim\limits_{k \to \infty} \alpha_k = \log_2
\lsr(A_0, A_1) \quad \mbox{and} \quad  \beta = \lim\limits_{k
\to \infty} \beta_k = \log_2 \jsr(A_0, A_1)  ,
\end{equation}
where the matrices $A_0, A_1$ are defined in Appendix \ref{ap-values}.
Moreover, there are positive constants $C_1, C_2$ such that
\begin{equation}\label{const}
C_1 \le \
\min\limits_{2^{k-1} <  n  \le  2^k} u_n n^{-\alpha } \quad \mbox{and} \quad C_1 \le
\max\limits_{2^{k-1} <  n  \le  2^k} u_n
n^{-\beta } \le \ C_2
\end{equation}
for all $k \in \n$.
\end{theo}
The proof of this theorem is based on the following auxiliary result
taken from~\cite{protasov3}. For a given set of indices $\{i_1 , \ldots
, i_p\}\subset \{1, \ldots , d\},\ p \ge 1$ we call the subspace $L_{i_1 ,
\ldots , i_p} = \{x \in \re^d,\ x_{i_1}= \cdots = x_{i_p} = 0\}$ a \emph{coordinate plane}.
\begin{propp}\label{p2}\cite{protasov3}
Let $A_0, A_1$ be matrices with a common invariant cone.
Then there is a positive constant $c_1$ such that
$$\max_{d_1, \ldots , d_k}\|A_{d_1}\cdots A_{ d_k}\|\ge \ c_1
\jsr^{ k}  \quad \mbox{and} \quad \min_{d_1, \ldots ,
d_k}\|A_{d_1}\cdots A_{ d_k}\|\ge \ c_1 \lsr^{ k}
, \quad k \in \n .$$
If, moreover, these matrices have no common invariant subspace
among the coordinate planes, then there is a positive constant
$c_2$ such that
$$\max_{d_1, \ldots , d_k}\|A_{d_1}\cdots A_{ d_k}\|\le \ c_2
\jsr^{ k}  , \qquad k \in \n .$$
\end{propp}
{\it Proof of Theorem~\ref{th2}}. The equalities in Eq. (\ref{ab}) follow
immediately from Corollary \ref{p1} and the definitions given on Eq. (\ref{jsrlsr}).
To prove the inequalities given at Eq. (\ref{const}) we
apply Proposition~\ref{p2} for our matrices $A_0, A_1$ \rj{that have} an
invariant cone $\re^{20}_+$. Theorem~\ref{th1} yields
$$ u_n n^{-\alpha } \asymp \|A_{d_1}\cdots A_{ d_k}\|
2^{-\alpha  k} = \|A_{d_1}\cdots A_{ d_k}\|\lsr
^{- k}.$$
Taking the minimum over $n = 2^{k-1}+1, \ldots , 2^k$ and invoking
Proposition~\ref{p2} we conclude that $\min\limits_{2^{k-1} <
n  \le  2^k} u_n n^{-\alpha } \ge  C_1$. The same with
the inequality $\max\limits_{2^{k-1} <  n  \le  2^k}
u_n n^{-\beta } \ge  C_1$. To prove the upper bound
in~ Eq. (\ref{const}) we note that the matrices $A_0, A_1$ have no
common invariant subspaces among the coordinate planes
\rj{(to see this observe, for instance, that $(A_0+A_1)^5$ has no zero entry).} {\hfill $\Box$}\\
\begin{corr}\label{c3}
There are positive constants $C_1, C_2$ such that
$$C_1n^{\alpha}  \le \ u_n \le  C_2n^{\beta}  , \ n \in \n.$$
\end{corr}
%
%
%
In the next section we will see that $\alpha < \beta.$  In particular, the sequence $u_n$
does not have a constant rate of growth, and the value $\frac{\log
 u_n}{\log  n}$ does not converge as $n \to \infty$. This was
already noted by Cassaigne in~\cite{cassaigne}. Nevertheless, it
appears that the value $\frac{\log u_n}{\log  n}$ actually has a
limit  as $n \to \infty$, not along all the natural numbers $n\in
\n$, but along a subsequence of $\n$ of density $1$. In other
terms, the sequence converges with probability $1$. The limit,
which differs from both $\alpha$ and $\beta$ can be expressed by
the so-called Lyapunov exponent $\bar \rho$ of the matrices $A_0,
A_1$. To show this we apply the following result proved by
Oseledets in 1968. For the sake of simplicity we formulate it for
two matrices, although it can be
easily generalized to any finite set of matrices.
\begin{theo}\label{th4}\cite{oseledets}
Let $A_0, A_1$ be arbitrary matrices and $d_1, d_2, \ldots
$ be a sequence of independent random variables that take values
$0$ and $1$ with equal probabilities $1/2$. Then the value
$\|A_{d_1}\cdots A_{d_k}\|^{1/k}$ converges to some number $\bar
\rho$ with probability $1$. This means that for any $\varepsilon >
0$ we have $ P \bigl( \bigl|  \|A_{d_1}\cdots
A_{d_k}\|^{1/k}   -  \bar \rho  \bigr| > \varepsilon
 \bigr)\to \ 0 $ as $ k \to \infty$.
\end{theo}
The limit $\bar \rho$ in Theorem~\ref{th4} is called the \emph{Lyapunov
exponent} of the set $\{A_0, A_1\}.$ This value is given by
the following formula:
\begin{equation}\label{lyap}
\bar \rho  (A_0, A_1) = \lim\limits_{k \to \infty}\
\Bigl(  \prod\limits_{d_1, \ldots , d_k} \|A_{d_1}\cdots
A_{d_k}\|^{1/k} \Bigr)^{{1}/{2^k}}
\end{equation}
(for the proof see, for instance, \cite{protasov4}). To understand
what this gives for the asymptotics of our sequence $u_n$ we
introduce some further notation. Let $\mathcal P$ be some property of
natural numbers. For a given $k \in \n$ we denote
 $$ P_k ({\mathcal{P}}) = \
2^{-(k-1)}{\rm Card} \bigl\{ n  \in \{2^{ k-1}+1, \ldots ,
2^{ k} \}, \quad n \,  \mbox{satisfies} \,  \mathcal P\ \bigr\}.$$
Thus, $P_k$ is the probability that the integer $n$ uniformly
distributed on the set $\{{2^{k-1}+1}, \ldots , 2^k\}$ satisfies
$\mathcal P$. Combining Proposition~\ref{p1} and Theorem~\ref{th4} we
obtain
\begin{theo}\label{th5}
There is a number $\sigma$ such that for any $\varepsilon > 0$ we
have $$P_k \Bigl( \Bigl|  \frac{\log  u_n}{\log  n} -
 \sigma\ \Bigr| > \varepsilon \Bigr)\to \ 0\qquad  \mbox{as} \ k \to \infty.$$
 Moreover, $ \sigma  =  \log_2  \bar \rho$, where $\bar \rho$ is the Lyapunov exponent of the matrices
 $\{A_0, A_1\}$ defined in Appendix \ref{ap-values}.
\end{theo}
Thus, for almost all numbers $n\in \n$ the number of overlap-free
words $u_n$ has the same exponent of growth $ \sigma  = \log_2
\bar \rho$. If positive $a$ and $b$  are large enough and $a< b $,
then for a number $n$ taken randomly from the segment $[a,b]$ the
value $ \frac{\log u_n}{\log  n} $ is close to $\sigma$. Let us
recall that a subset $\mathcal A \subset \n$ is said to have
density $1$ if $\frac{1}{n}{\rm Card} \bigl\{ r \le n,\ r \in
{\mathcal A} \bigr\}\to \ 1$ as $n \to \infty$. We say that a
sequence $f_n$ converges to a number $f$ along a set of density
$1$ if there is a set ${\mathcal A} \subset \n$ of density $1$
such that $ \lim\limits_{n \to \infty ,  n \in {\mathcal
A}} f_n = \ f$. Theorem~\ref{th5} yields\\

\begin{corr}\label{c4}
The value $ \frac{\log  u_n}{\log  n} $ converges to
$\sigma$ along a set of density $1$.
\end{corr}
{\it Proof.  } By Theorem~\ref{th5} there exists $k_1 \in \n$ such
that at least half of the natural numbers $n \le 2^{ k_1}$ satisfy the
inequality $\bigl|  \frac{\log  u_n}{\log  n}  -
  \sigma\ \bigr| <  \frac12$. Denote the set of such numbers $n \le 2^{ k_1}$
 by $\mathcal A_{ 1}$. Further, there exists $k_2 > k_1$ such that
at least $\frac{3}{4}$ of the numbers $n \in \{2^{ k_1}+1, \ldots
, 2^{ k_2}\}$ satisfy the inequality $\bigl| \frac{\log  u_n}{\log
n}  -   \sigma\ \bigr| <  \frac{1}{4}$. Denote the set of such
numbers by ${\mathcal A}_{ 2}$.
 The remainder is by induction: having a number $k_{ r-1}$ we take a number $k_{ r} > k_{ r-1}$
 such that
at least $\frac{2^{r}-1}{2^{r}}$ of the numbers $n \in \{2^{
k_{r-1}}+1, \ldots , 2^{ k_{r}}\}$ satisfy the inequality $\bigl|
\frac{\log  u_n}{\log  n}  -
  \sigma\ \bigr| <  \frac{1}{2^{r}}$ and denote the set of such numbers by ${\mathcal A}_{ r}$.
 The set $\mathcal A  =  \cup_{r \in \n} {\mathcal A}_{ r} $ has density $1$ and
 $\frac{\log  u_n}{\log  n}$ tends to $\sigma$ along $\mathcal A$ as $n \to \infty$.
{\hfill $\Box$}

\section{Estimations of the exponents}\label{section:numerical}

Theorems~\ref{th1} and~\ref{th5} reduce the problem of estimating the exponents of growth of $u_n$ to
computing joint spectral characteristics of the matrices $A_0$ and $A_1$. In order to estimate the joint spectral
radius we use a modified version of the ``ellipsoidal norm algorithm'' \cite{blondel-ellipsoid}. For the
lower spectral radius and for the Lyapunov exponent we present new algorithms, which seem to be relatively
efficient for nonnegative matrices.  The results we obtain can be summarized in the following theorem:\\

\begin{theo}\label{th-results}
\begin{equation}
\begin{array}{ccccc}
1.2690 & <&  \alpha & < & 1.2736\\
1.3322 & < & \beta & < & 1.3326\\
1.3005 &<& \sigma &<& 1.3098
\end{array}
\end{equation}
\end{theo}

In this section we also make (and give arguments for) the following conjecture:\\

\begin{conjj}\label{con1}
$$\beta =  \log_2  \sqrt{\rho (A_0A_1)}=1.3322\dots .$$
\end{conjj}

\begin{subsection}{ Estimation of $\beta$ and the joint spectral radius}
By Theorem~\ref{th2} to estimate the exponent~$\beta$ one needs
to estimate the joint spectral radius of the set $\{A_0, A_1\}$.
A lower bound for $\hat \rho$ can be obtained by applying inequality~(\ref{rho}).
Taking $k = 2$ and $d_1 = 0, d_2 = 1$ we get
\begin{equation}\label{lb-jsr}
\hat \rho \quad \geq \quad \bigl[\rho(A_0A_1)\bigr]^{1/2}=2.5179,
\end{equation}
and so $\beta \, > \, \log_2 2.5179 \, > \, 1.3322 $ (this lower bound was already found in \cite{cassaigne}).\\
Upper bounds for the joint spectral radius of sets of matrices
$\Sigma = \{A_{1}, \ldots , A_{m}\}$ are usually derived from the following simple inequality
\begin{equation}\label{ub-jsr}
\hat \rho \quad \leq \quad \max_{d_1, \ldots , d_k \, \in \, \{1, \ldots , m\}}
\|A_{d_1}\cdots A_{d_k}\|^{1/k},
\end{equation}
which holds for every $k \ge 1$ and converges to $\hat \rho\, $ as $\, k \to \infty$. This, at least
theoretically, gives arbitrarily sharp estimations for $\hat \rho$.
However, in our case, due to the size of the matrices $A_0, A_1$, this method leads to  computations
that are too expensive even for relatively small values of $k$.
Faster convergence can be achieved by finding an appropriate norm.
To do this we use the so-called ellipsoidal norm: $||A||_P=\max_{x}{\sqrt{\frac{x^TA^TPAx}{x^TAx}}}, $
where $P$ is a positive definite
matrix. This is the matrix norm induced by the vector norm $|x|_P=(x^TPx)^{1/2}.$
The crucial idea is that the optimal $P$, for which
the right hand side in~(\ref{ub-jsr}) for $k = 1$ is minimal, can be found by solving a simple
semidefinite programming problem.  This algorithm can be
iterated using the  relation $\rho(\Sigma^k)=\rho(\Sigma)^k$.
In the sequel we denote $\, \Sigma^k \, = \,
\bigl\{\, A_{d_1}\cdots A_{d_k}, 1\leq d_i\leq m,
i \, = \, 1, \ldots , k\, \bigr\}$. Thus
one can consider the set $\Sigma^k$ as a new set of matrices, and approximate its
joint spectral radius with the best possible ellipsoidal norm.  In Appendix \ref{ap-P} we give
an ellipsoidal norm such that each matrix in $\Sigma^{14}$ has a norm smaller than $2.5186^{14}.$
This implies that $\hat \rho \le 2.5186$, which gives
$\beta < 1.3326$. Combining this with the inequality $\beta > 1.3322$ we complete the proof
of the bounds for $\beta$ in Theorem~\ref{th-results}.\\
We have not been able to improve the lower bound of Eq. (\ref{lb-jsr}).  However, the upper bound
we obtain is very close to this lower bound, and the upper bounds obtained with an ellipsoidal norm
for $\Sigma^k$ get closer and closer to this value when $k$ increases.  Moreover, it has already been
observed that for many sets of matrices for which the joint spectral radius is known exactly, and in
particular matrices with nonnegative integer entries, there always is a  product that achieves the joint
spectral radius, i.e., a product $A\in \Sigma^t$ such that
$\hat \rho=\rho(A)^{(1/t)}$ \cite{jungers-blondel-finiteness}. For these reasons, we conjecture that
the exponent $\beta$ is actually equal to the lower bound.\end{subsection}

\begin{subsection}{Estimation of $\alpha$ and the lower spectral radius}
An upper bound for $\check \rho(A_0, A_1)$ can be obtained using Eq.(\ref{rho})
for $k = 1$ and $d_1 = 0$. We have
\begin{equation}
\alpha = \log_2(\check\rho)\leq \log_2(A_0)=1.276...
\end{equation}
This bound for $\alpha$ was first derived in~\cite{cassaigne}.
This bound is however not optimal. Taking the product
$A_1^{10}A_0$ (i.e., $k=11$ in inequality~(\ref{rho})),
we get a better estimate:
\begin{equation}\label{ub-lsr}
\alpha  \leq \log_2\, \bigl[ (\rho(A_1^{10}A_0)^{1/11}\bigr] =1.2735...
\end{equation}
One can verify numerically that this product gives the best possible upper bound among all the
matrix products of length $k \le 14$.

We now estimate $\alpha$ from below.  The problem of approximating the lower spectral radius is
NP-hard \cite{tsitsiklis97lyapunov} and to the best of our knowledge, no algorithm is
known to compute $\check \rho$. Here we propose two new algorithms.
We first consider nonnegative matrices.
As we observed above, for any $k$ we have $\, \check \rho (\Sigma^k) = \check \rho^k (\Sigma)$.
Without loss of generality it can be assumed that the matrices of the
set $\Sigma$ do not have a common zero column. Otherwise, by suppressing this column
and the corresponding row we obtain a set of matrices of smaller dimension with
the same lower spectral radius. The vector of ones is denoted by $\vecte$.\\

\begin{theo}\label{th7}
Let $\Sigma$ be a set of nonnegative matrices that do not have any common zero column. If for some
$r\in \re^+, s\leq t\in \n,$
there exists $x\in \mathbb{R}^d$ satisfying the following system of linear inequalities
\begin{equation}\label{progr1.lsr}
\begin{array}{lcl}
B(Ax-rx) & \geq &  0 , \quad  \forall B\in\Sigma^s,A\in\Sigma^t,\\
x & \geq &0, \quad (x, \vecte) = 1,
\end{array}
\end{equation}
then $\check\rho(\Sigma) \geq  r^{1/t}.$
\end{theo}

{\it Proof.  }
Let $x$ be a solution
of~(\ref{progr1.lsr}).  Let us consider a product of matrices $A_k\dots A_1 \in \Sigma^{kt}:A_i\in \Sigma^t.$
We show by induction
 in $k$ that $A_k\dots A_1 x \geq r^{k-1}A_kx:$ For $k=2,$ we have $A_2(A_1x-rx)=CB(A_1x-rx)\geq 0,$
 with $B\in\Sigma^s, C\in \Sigma^{t-s}.$ For $k>2$ we have
 $A_k\dots A_1 x = A_k A_{k-1}\dots A_1 x \geq r^{k-2} A_k A_{k-1}x \geq r^{k-1} A_k x.$
 In the last inequality the case for $k=2$ was reused.
 Hence, $||A_k \dots A_1||=\vecte^TA_k \dots A_1 \vecte \geq r^{k-1} \vecte^TA_kx\geq r^kC,$
 where $C=(\min_{k}\vecte^TA_kx)/r>0.$ The last inequality holds because $A_kx=0,$ together with the
 first inequality in (\ref{progr1.lsr}), imply that $-rBx=0$ for all $B\in \Sigma^s,$ which implies that
 all $B\in \Sigma^s$ have a common zero column.  This is in contradiction with our assumption because the
 matrices in $\Sigma^s$ share a common zero column if and only if the matrices in $\Sigma$ do. {\hfill $\Box$}\\
We were able to find a solution to the linear programming problem~(\ref{progr1.lsr}) with $r=2.41,\ t=16, s=6.$
Hence we get
the following lower bound: $\, \alpha \, \geq \log{r}> 1.2690.$  The corresponding vector $x$ is given in
Appendix \ref{ap-x}. This completes the proof of Theorem~\ref{th-results}.

Theorem~\ref{th7} handles nonnegative matrices, and we propose now
a way to generalize this result to arbitrary real matrices.  The
idea is to lift the matrices to a larger vector space, so that all the matrices share an invariant cone.
This kind of lifting is rather classical and is known under
several names in the literature as for instance semidefinite
lifting or symmetric algebras \cite{blondel-kron, protasov2,parrilo-jadbabaie}.  The idea is to consider
the matrices $A_i\in \Sigma$ as linear operators acting on
the cone of positive semidefinite matrices $S$ as
$S\rightarrow A_i^TSA_i.$  It is not difficult to prove that the
lower spectral radius of this new set of linear operators is equal
to $\check \rho (\Sigma)^2.$
We use the notation $A \succeq B$ to denote that the matrix $A-B$ is positive semidefinite.  Recall that
$A\succeq 0 \Leftrightarrow \forall y, y^TAy\geq 0.$
\begin{theo}\label{th8}
Let $\Sigma$ be a set of matrices in $\re^{d \times d}$ and $s \leq t\in\n.$
Suppose that there are
 $r >0$ and a symmetric matrix $S \succeq 0$ such that
\begin{equation}\label{progr2.lsr}
\begin{array}{ll}
B^*(A^*SA -rS)B  \succeq 0 & \forall A \in \Sigma^{ t},B\in \Sigma^{s}\\
S\succ 0 \end{array}
\end{equation}
then $\check\rho(\Sigma)\geq r^{1/2t}.$
\end{theo}

{\it Proof.  } The proof is formally similar to the previous one:  Let $ S $ be a solution
 of~(\ref{progr2.lsr}).  We denote $M_k$ the product $A_1 \dots A_k.$  It is easy to show by induction
 that $M_k^*SM_k   \succeq r^{k-1}(A_{k}^*SA_{k} ).$  This is obvious for $k=2$
 for similar reasons as in the previous theorem, and for $k>2,$ if, by induction,
 $$\forall y,\quad  y^*M_{k-1}^*SM_{k-1}y\geq r^{k-2} y^*A_{k-1}^*SA_{k-1}y,$$ then, with $y=A_kx,$
 for all $x,$  $$x^*M_k^*SM_kx\geq r^{k-2} x^*A_{k}^*A_{k-1}^*SA_{k-1}A_{k}x\geq r^{k-1}x^*A_{k}^*SA_{k}x.$$
Thus,
$$\sup{\left\{\frac{x^*M_k^*SM_kx}{x^*Sx}\right\}}\geq  r^{k-1}
\sup{ \left\{\frac{ x^*A_{k}^*SA_{k}x}{x^*Sx}\right\}}.
$$
Finally, $||M_k||_S\geq r^{k/2} C,$ where $C$ is a constant.
 %
%
{\hfill $\Box$}\\
For a given $r >0$ the existence of a solution $S$ can be
established by solving the semidefinite programming
problem~(\ref{progr2.lsr}), and the optimal $r$ can be found by
bisection in logarithmic time.
\end{subsection}

\begin{subsection}{Estimation of  $\sigma$ and the Lyapunov exponent}
The exponent of the average growth $ \sigma  $ is obviously between
$\alpha$ and $\beta$, so $ 1.2690   <  \sigma   <  1.3326$.  To get better bounds we need to
estimate the Lyapunov exponent $ \bar \rho $ of the matrices $A_0, A_1$. The first upper
bound can be given by the so-called 1-radius $\rho_1$:
$$
\rho_1 = \lim\limits_{k \to \infty} \Bigl(  2^{-k} \sum\limits_{d_1 , \ldots , d_k}
 \|A_{d_1}\cdots A_{d_k}\| \Bigr)^{1/k}.
$$
For matrices with a common invariant cone we have $ \rho_1  =  \frac12 \rho  (A_0 + A_1)$
\cite{protasov3}. Therefore, in our case $ \rho_1   =  \frac12 \rho  (A_0 + A_1)  =  2.479...$.
This exponent was first computed in~\cite{cassaigne}, where it was shown that the value
$\sum_{j=0}^{n-1} u_j $  is equivalent to $n^{ \eta}$, where $ \eta  =
1+\log_2  \rho_1  =  2.310...$.
It follows immediately from the inequality between the arithmetic mean and the geometric mean that
$ \bar \rho  \le  \rho_1  $. Thus, $ \sigma  \le  \eta$. In fact, as we
show below, $ \sigma$ is strictly smaller than $\eta$. We are not aware of any
approximation algorithm for the Lyapunov exponent, except by application of Definition~(\ref{lyap}).
It is easily seen that for any $k$ the value $r_k  =  \bigl(  \prod\limits_{d_1, \ldots , d_k}
\|A_{d_1}\cdots A_{d_k}\| \Bigr)^{\frac{1}{k2^k}}$ gives an upper bound for $\bar \rho$,
that is $ \bar \rho  \le r_k $ for any $k \in \n$. Since $r_k \to \bar \rho$ as $k \to \infty$,
we see that this estimation can be arbitrarily sharp for large $k$.
But for the dimension $20$ this leads to extensive numerical computations. For example,
for the norm $\|\cdot \|_1$ we have $r_{20} = 2.4865$, which is even larger  than $ \rho_1$.
In order to obtain a better bound for $\bar \rho$ we state the following results. For any $k  $ and
$ x \in \re^d  $ we denote
$ p_k(x) =  \bigl(\prod\limits_{d_1, \ldots , d_k} |A_{d_1}\cdots A_{d_k}x| \Bigr)^{\frac{1}{2^k}}$
and $ m_k  =  \sup\limits_{x \ge 0, |x| = 1} p_k(x)$.\\

\begin{propp}\label{p3}
Let $A_0, A_1$ be nonnegative matrices in $\re^d$.
Then for any norm $|\cdot |$ and for any $k \ge 1$ we have $ \bar \rho  \le  (m_k)^{1/k}$.
\end{propp}

{\it Proof.  } For any $x \ge 0$ and for any $n,k \in \n$ we have $ p_{k+n}(x)  \le  m_k  p_n(x)$.
Therefore, $ p_{tk}(x)  \le  (m_k)^t$. Whence,
$\lim\limits_{t \to \infty}\bigl[ p_{tk}(x)\bigr]^{1/tk} \le  (m_k^t)^{1/tk}  =  (m_k)^{1/k}$.
On the other hand, by Corollary~\ref{c1} for $x > 0$ we have $r_n \asymp [p_n(x)]^{1/n}$, and
consequently $ \lim\limits_{t \to \infty}\bigl[ p_{tk}(x)\bigr]^{1/tk} \to  \bar \rho$ as
$t \to \infty$. Thus, $ \bar \rho  \le  (m_k)^{1/k}$.
{\hfill $\Box$}\\

\begin{propp}\label{p4}
Let $A_0, A_1$ be nonnegative matrices in $\re^d$ that do not have common invariant subspaces
among the coordinate planes. If $ \check \rho  <  \hat \rho$, then
$ \bar \rho  <   \rho_1$.
\end{propp}

{\it Proof.  }Let $v_*$ be the eigenvector of the matrix $\frac12\bigl(A_0^* + A_1^* \bigr)$
corresponding to its  largest eigenvalue $ \rho_1$. Since the matrices have no
common invariant coordinate planes, it follows that $v_* > 0$.
Consider the norm $|x| = (x, v_*)$ on $\re^d_+$. Take some
$k \ge 1$ and ${y \in \re^d_+, |y| = (y, v_*) = 1}$, such that $p_k(y) = m_k.$
We have
$$
m_k = p_k(y) \le 2^{-k}\sum\limits_{d_1, \ldots , d_k}| A_{d_1}\cdots A_{d_k} y | = \
2^{-k}\sum\limits_{d_1, \ldots , d_k}\bigl( A_{d_1}\cdots A_{d_k} y ,  v_* \bigr)
$$
$$
 = \
\Bigl( y  ,  2^{-k} \bigl(  A_0^* + A_1^*  \bigr)^k v_* \Bigr)= \
\rho_1^k\bigl( y ,  v_*\bigr) = \rho_1^k.
$$
Thus, $m_k \le \rho_1^k$, and the equality is possible only if all $2^k$ values
$| A_{d_1}\cdots A_{d_k} y |$ are equal.  Since $\check\rho<\hat\rho,$ there must be a $k$ such that the
inequality is strict. Thus, $m_k < \rho_1^k$ for some $k$, and by Proposition~\ref{p3} we have
$ \bar \rho  \le  (m_k)^{1/k}
<  \rho_1$.{\hfill $\Box$}\\

We are now able to estimate $\bar \rho$ for the matrices $A_0, A_1.$ For the norm $|x| = (x, v_*)$ used
in the proof of Proposition~\ref{p4} the value
$ -\frac{1}{k} \log_2  m_k$ can be found as the solution of the following
convex minimization problem with linear constraints:
\begin{equation}\label{ln}
\begin{array}{lc}
\min &-\frac{1}{k 2^{ k}\ln 2} \sum\limits_{d_1 , \ldots , d_k  \in  \{0,1\}} \ln  \Bigl(
 x  ,  A_{d_1}^*\cdots A_{d_k}^* v_* \Bigr)\\
\mbox{s.t. }&  x \ge 0 , \quad (x ,  v_*) =  1.
\end{array}
\end{equation}
The optimal value of this optimization problem is equal to $-\frac{1}{k} \log_2  m_k$,
which gives un upper bound for $ \sigma  =  \log_2  \bar \rho$ (Proposition~\ref{p3}).
Solving this problem for $k = 12$ we obtain $ \sigma  \le  1.3098$. We finally provide a theorem that
allows us to derive a lower bound on $\sigma.$  The idea is identical to the one used in Theorem~\ref{th7},
but transposed to the Lyapunov exponent.\\

\begin{theo}\label{th-lb-lyap}
Let $\Sigma$ be a set of nonnegative matrices that do not have any common zero column.
If for some  $r_i\in \re_+, s\leq t\in \n,$
there exists $x\in \mathbb{R}_+^{d}$ satisfying the following system of linear inequalities
\begin{equation}\label{progr1.lyap}
\begin{array}{lcl}
B(A_ix-r_ix) & \geq &  0 , \quad  \forall B\in\Sigma^s,A_i\in\Sigma^t,\\
x & \geq &0, \quad (x, \vecte) = 1,
\end{array}
\end{equation}
then $\bar\rho(\Sigma) \geq  \prod_i{r_i}^{1/(t2^t)}.$
\end{theo}

The proof is similar to the proof of Theorem \ref{th7} and is left
to the reader.  Also, a similar theorem can be stated for general
matrices (with negative entries), but involving linear matrix
inequalities. Due to the number of different variables $r_i,$ one
cannot hope to find the optimal $x$ with SDP and bisection
techniques.  However, by using the vector $x$ computed for
approximating the lower spectral radius (given in Appendix
\ref{ap-x}), with the values $s=8,t=16$ for the parameters, one
gets a good lower bound for $\sigma:$ $ \sigma  \ge  1.3005$.

\end{subsection}

\section{Conclusions}
The goal of this paper is to precisely characterize the asymptotic rate of growth of the number of
overlap-free words.  Based on Cassaigne's description of these words with products of matrices,
we first prove that these matrices can be simplified, by decreasing the state space dimension
from $30$ to $20.$  This improvement is not only useful for numerical computations, but allows to
characterize the overlap-free words that ``count'' for the asymptotics: we call these words
\emph{purely overlap free,} as they can be expressed iteratively as the image of shorter purely overlap
free words.\\
 We have then proved that the lower and upper exponents $\alpha$ and $\beta$ defined by Cassaigne are
 effectively reached for an infinite number of lengths, and we have characterized them respectively as
 the logarithms of the \emph{lower spectral radius} and the \emph{joint spectral radius} of the simplified
 matrices that we constructed.  This characterization, combined with new algorithms that we propose to
 approximate the lower spectral radius, allow us to compute them within $0.4\%$. The algorithms we propose
 can of course be used to reach any degree of accuracy for $\beta$ (this seems also to be the case for
 $\alpha$ and $\sigma,$ but no theoretical result is known for the approximation of the lower spectral radius).
 The computational results we report in this paper have all been obtained in a few minutes of computation
 time on a standard PC desktop and can  therefore easily be improved.\\
Finally we have shown  that for almost all values of $n$, the number of overlap-free words of length $n$ do
not grow as $n^\alpha,$ nor as $n^\beta,$ but in an intermediary way as $n^\sigma,$ and we have provided
sharp bounds for this value of $\sigma.$\\
This work opens obvious questions: Can joint spectral
characteristics be used to describe the rate of growth of other
languages, such as for instance the more general repetition free
languages~? The generalization does not seem to be straightforward
for several reasons: first, the somewhat technical proofs of the
links between $u_n$ and the norm of a corresponding matrix product
take into account the very structure of these particular matrices,
and second, it is known that a bifurcation occurs for the growth
of repetition-free words: for some members of this class of
languages the growth is polynomial, as for overlap-free words, but
for some others the growth is exponential \cite{karhumaki-shallit}, and one could wonder
how the joint spectral characteristics developed in this paper
could represent both kinds of growth.

\section*{Acknowledgment}\label{ac}
We would like to thank Prof. Stephen Boyd (Stanford University),
Yuri Nesterov, and Fran\c{c}ois Glineur (Universit\'e catholique
de Louvain) for their helpful suggestions on semi-definite programming
techniques. This research was carried out during the visit of the
second author to the Universit\'e catholique de Louvain
(Louvain-la-Neuve, Belgium). That author is grateful to the
university for its hospitality.

\newpage
\appendix

\section{Numerical values}\label{ap-values}
We introduce the following auxiliary matrices. For the  sake of
simplicity our notation do not follow exactly those of
\cite{cassaigne}.
$$D_1=\begin{pmatrix}0&0&0&0&0&0&0&1&2&1 \\0&0&1&1&0&1&1&0&0&0\\
0&0&0&0&0&0&0&1&1&0\\0&0&0&0&0&0&0&0&0&0\\1&2&0&0&1&0&0&0&0&0\\0&0&1&0&0&1&0&0&0&0\\
0&0&0&0&0&0&0&0&0&0\\0&0&0&0&0&0&0&1&0&0\\0&0&0&0&0&0&0&0&0&0\\0&0&0&0&0&0&0&0&0&0
\end{pmatrix},
B_1=\begin{pmatrix}0&0&0&0&0&0&0&1&2&1\\0&0&0&0&0&0&0&0&0&0\\0&0&0&0&0&1&1&0&0&0\\
0&0&1&1&0&0&0&0&0&0\\0&0&0&0&0&0&0&0&0&0\\0&0&0&0&0&0&0&0&0&0\\0&0&0&0&0&0&0&0&0&0\\
0&0&0&0&1&0&0&0&0&0\\0&1&0&0&0&0&0&0&0&0\\1&0&0&0&0&0&0&0&0&0\\
\end{pmatrix}, C_1=\begin{pmatrix}0&0&0&0&0&0&0&2&4&2\\0&0&1&1&0&1&1&0&0&0\\
0&0&0&0&0&1&1&1&1&0\\0&0&1&1&0&0&0&0&0&0\\0&0&0&0&0&0&0&0&0&0\\0&1&0&0&1&0&0&0&0&0\\
1&1&0&0&0&0&0&0&0&0\\0&0&0&0&0&2&0&0&0&0\\0&0&1&0&0&0&0&0&0&0\\0&0&0&0&0&0&0&0&0&0\\
\end{pmatrix},$$
$$B_2=\begin{pmatrix}0&0&0&0&0\\0&0&0&0&0\\0&0&0&1&1\\0&1&1&0&0\\0&0&0&0&0\\0&0&0&0&0\\
0&0&0&0&0\\0&0&0&0&0\\1&0&0&0&0\\0&0&0&0&0
\end{pmatrix},
C_2=\begin{pmatrix}0&0&0&0&0\\
0&0&0&0&0\\0&0&0&0&0\\0&0&0&0&0\\0&0&0&0&0\\0&0&0&0&0\\0&0&0&0&0\\0&0&0&2&0\\0&1&0&0&0\\0&0&0&0&0
\end{pmatrix}, C_4=\begin{pmatrix}0&1&1&1&1\\0&0&0&1&1\\0&1&1&0&0\\1&0&0&0&0\\1&0&0&0&0
\end{pmatrix}.$$
Now, defining \begin{equation}\label{F0F1} F_0=\begin{pmatrix} C_1&0_{10\times10}&
\vline&C_2&0_{10\times 5}\\D_1&B_1&\vline&0_{10\times5}&B_2\\\hline 0_{5\times 10}&
0_{5\times 10}&\vline&C_4&0_{5\times 5}\\ 0_{5\times 10}&0_{5\times10}&\vline&0_{5\times 5}&0_{5\times 5}
\end{pmatrix},F_1=\begin{pmatrix} D_1&B_1&\vline&0_{10\times 5}&B_2\\0_{10\times 10}&C_1&\vline&
0_{10\times5}&C_2\\\hline 0_{5\times 10}&0_{5\times 10}&\vline&0_{5\times 5}&0_{5\times 5}\\
0_{5\times 10}&0_{5\times10}&\vline&0_{5\times 5}&C_4 \end{pmatrix},\end{equation}
$$w=(1,2,2,2,1,2,2,1,2,1,0_{1\times20})^T,$$
$$ y_3=(2,0,2,0,2,0,0,0,0,2,0,2,2,0,0,0,2,2,0,0,0,0,0,0,0,0,0,0,0,0)^T, $$
$$y_4=(0,2,2,0,0,0,2,2,0,0,0,0,0,0,0,2,2,0,2,2,2,0,2,0,2,0,0,0,0,0)^T,$$
$$y_5=(2,2,0,2,2,2,0,2,0,2,0,0,0,0,0,4,2,0,2,0,2,2,0,2,0,0,2,0,0,0)^T,$$
$$y_6=(4,2,0,2,0,2,2,0,2,0,0,2,0,0,0,4,2,2,2,4,2,0,0,2,2,0,0,0,0,0)^T,$$
$$y_7=(4,2,2,2,4,2,0,0,2,2,0,0,0,0,0,4,4,4,2,0,2,2,0,2,0,0,0,0,0,2)^T,$$
$$y_8=(4,4,4,2,0,2,2,0,2,0,0,0,0,0,2,6,4,4,2,4,2,0,4,2,2,0,0,0,0,0)^T,$$
$$y_9=(6,4,4,2,4,2,0,4,2,2,0,0,0,0,0,8,4,4,2,0,4,4,4,0,0,0,0,0,0,0)^T,$$
$$y_{10}=(8,4,4,2,0,4,4,4,0,0,0,0,0,0,0,8,4,6,4,8,2,0,4,2,4,0,0,0,0,0)^T,$$
$$y_{11}=(8,4,6,4,8,2,0,4,2,4,0,0,0,0,0,8,6,6,2,0,2,6,4,2,0,2,0,2,2,0)^T,$$
$$y_{12}=(8,6,6,2,0,2,6,4,2,0,2,0,2,2,0,10,6,4,4,8,2,0,4,2,4,0,0,0,0,0)^T,$$
$$y_{13}=(10,6,4,4,8,2,0,4,2,4,0,0,0,0,0,12,6,4,4,0,6,6,4,2,0,0,0,0,0,0)^T,$$
$$y_{14}=(12,6,4,4,0,6,6,4,2,0,0,0,0,0,0,10,6,8,6,12,4,0,0,4,4,0,0,0,0,0)^T,$$
$$y_{15}=(10,6,8,6,12,4,0,0,4,4,0,0,0,0,0,8,10,6,6,0,4,8,4,4,0,2,2,0,0,0)^T,$$
and introducing the recurrence relation
$$ y_{2n}=F_0y_n,\quad y_{2n+1}=F_1y_n, \quad n\geq 8$$
one has the relation \cite{cassaigne}:
\begin{equation}\label{formulas-cassaigne}u_{n+1} = w^T y_n. \end{equation}
We finally introduce two new matrices in $\re^{20 \times 20}$ that
rule the asymptotics of $u_n:$
\begin{equation}\label{A0A1}
A_0=\begin{pmatrix} C_1&0_{10\times10}\\D_1&B_1\end{pmatrix},A_1=\begin{pmatrix}
D_1&B_1\\0_{10\times 10}&C_1 \end{pmatrix}.
\end{equation}

\section{Proof of Theorem \ref{th1}}\label{ap-th1}
In this appendix we give a proof of Theorem \ref{th1}.  We first
find a common invariant cone $K$ for the matrices $A_0,A_1.$   We
then show that the products $F(n)=F_{d_1}\cdots F_{d_k}$ are
asymptotically equivalent to their corresponding product
$A(n)=A_{d_1}\cdots A_{d_k}.$  We then shed some light on the
vectors $z_n,$ the restriction of $y_n$ to $\re^{20}:$ their norms
can be considered as norms of the products $A(n).$  We finally
show that $\|A_{d_1}\cdots A_{d_k}\|$ is equivalent to
$\|A_{d_1}\cdots A_{d_{k-4}}\|.$
We end with the proof of Theorem \ref{th1} that puts all this together.\\
Let us first establish a special property of the matrices $A_0, A_1$.
Consider the following sets:
$$\begin{array}{l}
P = \Bigl\{x \in \re^{20},
x \ge 0,\quad  x_i >0  \mbox{ for all } i  \notin  \{5, 10, 17\} \Bigr\} , \\
{}\\ Q = \Bigl\{x \in \re^{20},   x \ge 0,\quad x_i >0  \mbox{
for all } i  \notin  \{7, 15, 20\} \Bigr\}.\end{array}$$
Let $S = P\cup Q$. For any $\varepsilon \ge 0$ let
$p_{\varepsilon} \in \re^{20},(p_{\varepsilon})_i =
-\varepsilon$ for $i \in \{5, 10, 17\}$ and $(p_{\varepsilon})_i =
1$ otherwise, let also $q_{\varepsilon} \in \re^{20},\
(q_{\varepsilon})_i = -\varepsilon$ for $i \in \{7, 15, 20\}$ and
$(q_{\varepsilon})_i = 1$ otherwise. This is easy to verify \rj{by direct calculation} that
there is $\varepsilon > 0$ such that $A_0p_{\varepsilon},A_0q_{\varepsilon} \in P$ and
$A_1q_{\varepsilon},A_1p_{\varepsilon} \in Q$ (this is the case for instance for $\varepsilon = 1/4$).
Take any such $\varepsilon$
and denote $ K  =  \{v + tp_{\varepsilon} +
sq_{\varepsilon},v \in \re^{20}_+, t,s \ge 0\}$. Clearly, $K$
is a convex closed pointed cone.\\

\begin{lema}\label{l2}
We have $A_iK \subset K$ and $A_iS \subset S $ for $ i = 0,1$.
Moreover, $  S \subset  {\rm int } K$.
\end{lema}

{\it Proof.  } Let $x\in K:\, x = v + tp_{\varepsilon} + sq_{\varepsilon},v
\ge 0, t,s \ge 0$. Since $A_iv \ge 0$, $A_ip_{\varepsilon} \ge 0$
and $A_iq_{\varepsilon} \ge 0$ we see that $A_i x \ge 0$. Thus,
$A_iK\subset K$.\\ Further, for an arbitrary $x \in S$ let $x_{\min}
= \min\limits_{x_i >0} x_i$. Assume $x \in P$ (the proof for the
case $x \in Q$ is literally the same). Since $x \ge x_{\min}p_0$
we have $ A_0x  \ge x_{\min}A_0p_0  \ge  x_{\min}p_0 \in P $ and $
A_1x  \ge
 x_{\min}A_1p_0  \ge  x_{\min}q_0 \in Q$. Therefore $ A_ix
\in S$.\\ Finally, for any $x \in S$ and $y\in \re^{20}$ such that
$|y|_\infty<\min{\{\frac12  \, \varepsilon \, x_{\min}\, ,\,
\frac{1}{2}\varepsilon\}}$ we have $x+y\in K.$ This proves that $x
\in {\rm int} K.$
{\hfill $\Box$}\\

\begin{corr}\label{c2}
For any $x \in S$ and for any sequence  $d_1, \ldots , d_k$ we
have $A_{d_1}\cdots A_{d_k}x \in  {\rm int} K$.\end{corr}

\begin{lema}\label{l3}
Suppose $n\geq 1$ and consider   the binary expansion $d_k\dots
d_1$ of the number $n-1$. We define  $A(n)=A_{d_1}\cdots A_{d_k}$
and similarly for $F(n)$. Then, for any matrix norm one has:
$$\|F(n)\|  \asymp  \|A(n)\|.$$\end{lema}

{\it Proof.  }Since all matrix norms are equivalent, it suffices to
consider any norm. Obviously $\|F_{d_1}\cdots F_{d_k}\|_{1} \ge
\|A_{d_1}\cdots A_{d_k}\|_{1}$ (\rj{because the matrices $A_i$ are submatrices of $F_i$}), hence it remains
to prove the
opposite inequality: there is a positive constant $C$ such that
$ \|A_{d_1}\cdots A_{d_k}\|_{1}  \ge  C\|F_{d_1}\cdots
F_{d_k}\|_{1}$ for all $k \in \n$ and $d_1, \ldots , d_k$.
We consider the case where $d_k=1;$ the proof for the other case is similar. Let $m \le k-1$ be the biggest
number such that $d_m =0$. If the sequence has no zero, we fix $m=0.$ \rj{Let $A_i,$ $H_i,$ and $R_i$ denote
respectively the upper left, the upper right and the lower right corners of the matrix $F_i$ in
bloc representation \ref{F0F1}}. Then the product $F_{d_1}\dots F_{d_k}$ has the following form: the left
upper block
is $A_{d_1}\cdots A_{d_k}$, the left lower block is zero, the
right lower block is $R_{d_1}\cdots R_{d_k}$, finally, the right
upper block is
\begin{equation}\label{summa1}
\sum\limits_{p=1}^{k}\quad \Bigl( \prod_{j =1}^{ p-1}A_{d_j}\Bigr)
 \cdot  H_{d_{p}} \cdot  \Bigl( \prod_{j =p+1}^{
k}R_{d_j}\Bigr).
\end{equation}
\rj{By convention,} the product over an empty set is one. Since $R_0R_1=0$, the
right lower block is zero and block (\ref{summa1}) becomes
$\sum\limits_{p=m}^{k}\left( \prod_{j=1}^{ p-1}A_{d_j}\right)
H_{d_{p}} R_{1}^{k-p}$, whose norm can be estimated from above as
\begin{equation}\label{summa2}
H \sum\limits_{p=m}^{k}\quad \Bigl\|  \prod_{j =1}^{ p-1}A_{d_j} \Bigr\|_1  \cdot  \Bigl\| R_{1}^{k-p}
\Bigr\|_1,
\end{equation}
where $H = \max \{\|H_0\|_1, \|H_1\|_1\}$. It was shown
in~\cite{cassaigne} that the sum of entries of the
matrix~$R_i^{l}$ does not exceed $ C 2^{l}$ for any $i = 0,1$
and $l \ge 1$, where $C >0$ is a constant. Hence $\|R_1^{k-p}\|_1
 \le  C 2^{k-p}$. Thus,
\begin{equation}\label{summa3}
\|F_{d_1}\cdots F_{d_k}\|_1 \le \|A_{d_1}\cdots A_{d_k}\|_1 \
+ HC \sum\limits_{p=m}^{k}2^{k-p} \Bigl\| \prod_{j =1}^{
p-1}A_{d_j}\Bigr\|_1.
\end{equation}
On the other hand, for any $p \ge m$ we have $|A_{d_1}\cdots
A_{d_k}\vecte|  =  |A_{d_1}\cdots A_{d_{p-1}}(A_{d_p}A_1^{k-p}\vecte)|$,
where~$\vecte$ is the vector of ones. By Corollary~\ref{c2}  the
vectors $A_{d_p}A_1^{r}\vecte$ belong to ${\rm int } K$ for all $r \in
\n$. Moreover, the vector $A_1^{r}\vecte/|A_1^{r}\vecte|$ converges to $v_1$
(the Perron-Frobenius eigenvector of~$A_1$) as $r \to \infty$ and
$v_1 \in S$, hence $v_1 \in {\rm int} K$ (Lemma~\ref{l2}).
Therefore there is a constant $C_1>0$ such that $\gamma
(A_{d_p}A_1^{r}\vecte) \ge C_1$ for all $r \in \n$. Applying now
Lemma~\ref{l1} for $x = A_{d_p}A_1^{k-p}\vecte$, we get
$$\|A_{d_1}\cdots A_{d_k}\| \ge \
 C_2 |A_{d_1}\cdots A_{d_{p-1}}(A_{d_p}A_1^{k-p}\vecte)|\
 $$
 $$
\ge C_2 C_1 \|A_{d_1}\cdots A_{d_{p-1}}\|\cdot
|A_{d_p}A_1^{k-p}\vecte| \ge C_3 \lambda^{k-p}\|A_{d_1}\cdots
A_{d_{p-1}}\|,
$$
where $\lambda = \rho(A_1)$ (by the same \rj{reasoning we have
$|A_{d_p}A_1^{r}\vecte| \geq \gamma(A_1^{r}\vecte)||A_{d_p}||\cdot |A_1^{r}\vecte| \ge C\lambda^{r}$,
indeed, $A_1^r \vecte\in \rm{int} K,$ and thus $||A^{d_p}A_1^{r}\vecte||\asymp ||A^{d_p}||\cdot
|A_1^{r}\vecte|$}). Thus, $\|A_{d_1}\cdots A_{d_{p-1}}\| \le
C_3^{-1}\lambda^{p-k}\|A_{d_1}\cdots A_{d_k}\|$ for all $p \le k$.
Substituting this in~(\ref{summa3}) and taking into account that
$ \frac{2}{\lambda}  <  1 $ (because $ \lambda  =
\rho(A_1)  >  2.42$) we take the sum of the geometrical
progression and get $\|F_{d_1}\cdots F_{d_k}\|_1  \le  C_4
\|A_{d_1}\cdots A_{d_k}\|_1$, where $C_4$ is some constant. This
concludes the proof.
{\hfill $\Box$}\\
\begin{lema}\label{l4}
For any $n$ we have $ u_{n+1} \le 2 u_n.$
\end{lema}
{\it Proof.  } If a word of length $n+1$ is overlap-free then so is its
prefix of length $n$. On the other hand, at most two
overlap-free words of length $n+1$ have the same prefix of length
$n$. {\hfill $\Box$}\\
\begin{lema}\label{l5}
Let the vectors $y_m\in \re^{30}$ be the solution of the recurrence equation \ref{eq-cassaigne},
and $z_m \in \re^{20}$ be the vector with the first $20$ entries of $y_n.$  We have $z_m \in S$
for each $m = 64, \ldots , 127$.
\end{lema}
{\it Proof.  } The proof is by direct calculation.
{\hfill $\Box$}\\
\begin{lema}\label{lem-k-4}
Suppose $n\in \n$ and $d_k \dots d_1$ is the binary expansion of
$n-1$; then $\|A(n)\| \asymp \|A'(n)\|$, where $A(n)=A_{d_1}\cdots
A_{d_{k}}$ and $A'(n)=A_{d_1}\cdots A_{d_{k-4}}$.
\end{lema}
{\it Proof.  }
The inequality $\bigl\|A_{d_1}\cdots A_{d_{k}}\bigr\|\leq C \bigl\|A_{d_1}\cdots A_{d_{k-4}}\bigr\|$
is obvious by submultiplicativity of the norm.  For the other direction, we have:
\begin{equation}\label{est1}
\bigl\|A_{d_1}\cdots A_{d_{k}}\bigr\|\asymp \
\bigl|A_{d_1}\cdots A_{d_{k}}\vecte\bigr| \asymp \
\bigl|A_{d_1}\cdots A_{d_{k-4}}(A_{d_{k-3}}\dots
A_{d_{k}}\vecte)\bigr|.
\end{equation}
Corollary~\ref{c2} yields $A_{d_{k-3}}\dots A_{d_{k}}\vecte
\in  {\rm int} K$ for all $d_{k-3},\dots, d_k \in \{0,1\}$.
Applying now Lemma~\ref{l1} we get $$ \delta =
\min\limits_{d_{k-3},\dots, d_k \in \{0,1\}} \gamma
(A_{d_{k-3}} \dots A_{d_{k}}\vecte) >  0.$$ Therefore, for some $C_1 > 0,$
\begin{equation}
\begin{tabular}{ccc}
$ C_1\bigl\|A_{d_1}\cdots
A_{d_{k-4}}\bigr\|$&$\le$&$ \delta  \bigl\|A_{d_1}\cdots
A_{d_{k-4}}\bigr\|\cdot \bigl|A_{d_{k-3}} A_{d_{k-1}}
A_{d_{k}}\vecte\bigr|$\\
&$ \le$&$\bigl|A_{d_1}\cdots A_{d_{k}}\vecte\bigr|.$
\end{tabular}
\end{equation}
Combining this with~(\ref{est1}) we get
$\bigl\|A_{d_1}\cdots A_{d_{k-4}}\bigr\|  \le  C_2
\bigl\|A_{d_1}\cdots A_{d_{k}}\bigr\|.$
{\hfill $\Box$}\\
We are now able to prove Theorem \ref{th1}:\\

{\it Proof of Theorem \ref{th1}. }Let $g$ be the vector of
$\re^{30}$, whose first $20$ entries are ones and the last $10$
entries are zeros. Let also $m = d_{k-3} + d_{k-2} 2 + 2^2 d_{k-1} + 2^3 d_k$.  Since $w\leq 2 g,$ we have
\begin{equation}\label{est0}
\begin{tabular}{ccccc}
$u_n$&$\le$&$2\bigl(y_{n-1}  ,  g \bigr)$&$=$&$ 2\bigl(F_{d_1}\cdots F_{d_{k-4}}y_{m} ,  g\bigr)$\\
&$\le$&$C_0\|F_{d_1}\cdots F_{d_{k-4}}\|$&$ \asymp $&$\|A_{d_1}\cdots A_{d_{k-4}}\|,$
\end{tabular}
\end{equation}
where $C_0$ does not depend on $n$ (the first two relations are
direct from  fundamental assertions \ref{formulas-cassaigne}, the
third relation comes from the fact that $y_m$ and $g$ are bounded,
and the last equivalence is by Lemma~\ref{l3}).  Combining Lemma
\ref{lem-k-4} and~(\ref{est0}) gives $u_n \le C_3
\bigl\|A_{d_1}\cdots A_{d_{k}}\bigr\|$.

Let us now prove the opposite inequality. Lemma~\ref{l4}, together with the fact that by construction,
the first ten entries of $y_n$ are equal to the entries $11,\dots, 20$ of $y_{n-1},$ imply that
$ u_n  \ge   \frac{1}{3}\bigl( u_n + u_{n+1}\bigr) \ge
\frac{1}{6}\bigl( y_{n-1} ,  g\bigr)$. Furthermore, for $n >
2^{ 7}$ we have $ \bigl( y_{n-1} ,  g\bigr)= \
\bigl(F_{d_1}\cdots F_{d_{k-7}}y_l ,  g\bigr)$, where $l =
\sum_{j=0}^6 d_{k-6+j} 2^{ j}$.  Thus,
\begin{equation}\label{est2}
u_n \ge \rj{\frac{1}{6}}\bigl(F_{d_1}\cdots
F_{d_{k-7}}y_l ,  g\bigr).
\end{equation}
On the other hand, defining $z_l\in \re^{20}$ as the vector with the first $20$ entries of $y_l:$
$\bigl(F_{d_1}\cdots F_{d_{k-7}}y_l ,
g\bigr)  \ge  \bigl(A_{d_1}\cdots A_{d_{k-7}}z_l ,
\vecte\bigr)  \rj{=}  \bigl|A_{d_1}\cdots A_{d_{k-7}}z_l\bigr|_1$. By
Lemma~\ref{l5} we have $z_l \in {\rm int} K$ for all $l \in
\{64, \ldots , 127\},$ and we can define ${ h  =   \min\limits_{64
\le l \le 127} \gamma (z_l) > 0}$ such that $\bigl|A_{d_1}\cdots
A_{d_{k-7}}z_l\bigr|_1  \ge  h C_4\|A_{d_1}\cdots
A_{d_{k-7}}\|$, where $ C_4  =   \min\limits_{64 \le l \le
127} |z_l|_1$. Combining this with~(\ref{est2}), we obtain
\begin{equation}\label{est3}
u_n \ge C_5  \|A_{d_1}\cdots A_{d_{k-7}}\|.
\end{equation}
Now, by submultiplicativity of the norm,
$$u_n \ge C_6  \|A_{d_1}\cdots A_{d_{k}}\|. $$
{\hfill $\Box$}\\

\section{The ellipsoidal norm}\label{ap-P}
Define 
$$P_1=
\begin{pmatrix}
   31.3  &  7.5 &   2.3  &  3.3  & -0.4 &  -0.3 &   0.3 &   0.4 &   3.7&0.3 \\
    7.5  & 57.7 &  10 &  6.3 &  18.4 &  35  & 16.3  & -5.8&   13.8&5\\
    2.3   &10 &  59.9  & 11.3 &   0.4 &  29.2  &  4.2 &  10.1 &   8.2&0.8\\
    3.3&    6.3 &  11.3  & 48.5 &   4.6 &  13.5  & 10.8&    2  &  6.9&1\\
   -0.4 &  18.4 &   0.4  &  4.6 &  36.4 &  23.5  & 22.6  &  4.4   & 8.9&-1.2\\
   -0.3 &  35 &  29.2  & 13.5 &  23.5 & 105.9  & 38.4 &   9.5&   33.7&6.1\\
    0.3 &  16.3 &   4.2  & 10.8 &  22.6   &38.4  & 59&    2.7 &  17.4&9.2\\
    0.4 &  -5.8 &  10.1  &  2 &   4.4  &  9.5  &  2.7  & 38.6  & 14.8&-1.7\\
    3.7 &  13.8 &   8.2  &  6.9 &   8.9  & 33.7  & 17.4 &  14.8   &57.5&8.6\\
    0.3 &   5 &   0.8  &  1 &  -1.2  &  6.1  &  9.2&   -1.7&    8.6&42.3\end{pmatrix},$$
 $$P_2=
\begin{pmatrix}
-10.4&   -1.7&  -18.1&   -0.4&   -5.8&   -5.1&   -4.9&   -0.8&   -2.7&   -0.9\\
-11.1  &-22.4 &  -8.2 & -14.7 &  -9.9 & -30.3&  -16.7 & -11.3&-16.9&   -6.6\\
 -2.2  &-16.4 & -15.8 &  -5 &  -8.5 &  -7.2&   -5.4 & -18.5&-3.5&   -3.4\\
 -0.2 & -13.6 &  -5.2 &  -9  &-10.7&  -14.6&   -9.2 &  -1.6&-11.3&   -1.1\\
 -4.6 & -17 & -13 &  -9.1 &  -0.6&  -11.2&  -23.9 &  -7& -12.1 &   0.3\\
  -5.9 & -26.4 & -27.4 & -17.4 & -31&  -37.6&  -28 &  -4.4&-27.3 &  -7.4\\
  -1.4 & -19.3 & -11.6  &-10.8 & -22.3&  -17.9&  -11.7 & -11.3&-12 &  -9.8\\
  -6.3 &   2.1 &   1.7  & -3.4 &   3.2&   -7.6 &   0.2 &  -5.2&-3.1 &  -1.4\\
  -7.4 & -15.9 &  -4.7 &  -6.7 & -12.2&  -17.3 & -11.6 &  -5.3&-6.8  & -1.6\\
   1.3 &  -5.7 &  -3.6 &  -3.2 &  -0.4&   -6.1 &  -9 &  -1.4&-6.9  &  0.4\\
   \end{pmatrix},$$

$$P_4=
\begin{pmatrix}
 29.1&  8.3 &  -1.6 &   4.8 &  -1.3&   -4.4 &   0.6 &   1.7&7.5  &  1.1\\
   8.3 &  47.3  & 13.6 &   2.8 &  11.7&   19.8 &  17.4 &   0.6&10 &  3.7\\
 -1.6 &  13.6  & 46.6  & 10.4 &   6.5&   24.9 &  11.8 &   6.5&12.5  &  1.4\\
  4.8 &   2.8 &  10.4  & 47.6 &   5.1&    8&   7.6 &   5.1&3.7  &  1.8\\
  -1.3 &  11.7 &   6.5 &   5.1 &  32.8&   19.5 &  19.4 &   7.6&6.7  & -0.2\\
  -4.4 &  19.8 &  24.9 &   8&  19.5&   64.8 &  16.2 &  11.4&13.8  &  6.8\\
 0.6 &  17.4 &  11.8  &  7.6 &  19.4&   16.2 &  56.7 &   7.6&12.2  &  6.5\\
  1.7 &   0.6 &   6.5  &  5.1 &   7.6 &  11.4 &   7.6 &  38.7&11.2  & -1\\
   7.5 &  10 &  12.5  &  3.7 &   6.7 &  13.8 &  12.2 &  11.2&55.6  &  4.2\\
  1.1 &   3.7&    1.4  &  1.8 &  -0.2  &  6.8   & 6.5 &  -1&4.2  & 43.8
  \end{pmatrix},
$$
$$
P=
\begin{pmatrix}
   P_1 &  P_2    \\
   P_2^T &  P_4  \end{pmatrix}.$$

Then one has the relations: $$ A^tPA-(2.5186)^{28}P\prec 0, \quad \forall A\in \Sigma^{14}.
$$
As explained in \cite{blondel-ellipsoid} this suffices to prove that $\rho(\Sigma)\leq 2.5186$.

\section{The vector $x$}\label{ap-x}

Define
$$
x=(    153,         0,    60,         0,    50,    56,    99,         0,    58,    1,    157, 81,
0,         113,         0,    72,    0,    99,         0,    0)^T.
$$
Then one has the relation
\begin{equation}
\begin{array}{lcl}
B(Ax-rx) & \geq &  0 , \quad  \forall B\in\Sigma^6,A\in\Sigma^{16},\\
x & \geq &0,
\end{array}
\end{equation}
with $r=2.41^{16}.$
this proves that $\check\rho(\Sigma) \geq  2.41.$

\end{document}